\documentclass[a4paper,english,preprint,floats]{revtex4-2}
\usepackage{helvet}

\usepackage[T1]{fontenc}
\usepackage[latin9]{inputenc}
\usepackage{babel}
\usepackage{array}
\usepackage{prettyref}
\usepackage{units}
\usepackage{amstext}
\usepackage{graphicx}
\usepackage[unicode=true,pdfusetitle,
 bookmarks=true,bookmarksnumbered=false,bookmarksopen=false,
 breaklinks=false,pdfborder={0 0 0},pdfborderstyle={},backref=false,colorlinks=false]
 {hyperref}

\makeatletter

\providecommand{\tabularnewline}{\\}

\usepackage{prettyref}
\newrefformat{fig}{\hyperref[#1]{Fig.~\ref*{#1}}}
\newrefformat{tab}{\hyperref[#1]{Table~\ref*{#1}}}

\usepackage{ifthen}
\newboolean{isarxiv}
\setboolean{isarxiv}{true}

\usepackage{lineno}
\ifthenelse{\boolean{isarxiv}}{}{\linenumbers}

\makeatother

\begin{document}
\title{Sub-realtime simulation of a neuronal network of natural density}
\author{Anno C. Kurth$^{1,2\ast}$, Johanna Senk$^{1}$, Dennis Terhorst$^{1}$,
Justin Finnerty$^{1}$ and Markus Diesmann$^{1,3,4}$}
\affiliation{$^{1}$Institute of Neuroscience and Medicine (INM-6) and Institute
for Advanced Simulation (IAS-6) and JARA-Institute Brain Structure-Function
Relationships (INM-10), Jülich Research Centre, Jülich, Germany~\\
$^{2}$RWTH Aachen University, Aachen, Germany~\\
$^{3}$Department of Psychiatry, Psychotherapy and Psychosomatics,
School of Medicine, RWTH Aachen University, Aachen, Germany~\\
$^{4}$Department of Physics, Faculty 1, RWTH Aachen University, Aachen,
Germany~\\
$^{\ast}$a.kurth@fz-juelich.de\vfill{}
~
}\begin{abstract}
Full scale simulations of neuronal network models of the brain are
challenging due to the high density of connections between neurons.
This contribution reports run times shorter than the simulated span
of biological time for a full scale model of the local cortical microcircuit
with explicit representation of synapses on a recent conventional
compute node. Realtime performance is relevant for robotics and closed-loop
applications while sub-realtime is desirable for the study of learning
and development in the brain, processes extending over hours and days
of biological time.\vfill{}
\newpage{}
\end{abstract}
\maketitle

\section*{Introduction }

The cortical neuronal network of mammals exhibits a two-fold universality:
basic characteristics of its architecture are conserved in evolution
from mouse to human as well as across brain areas. This has motivated
researchers to investigate models of the local cortical microcircuit,
the network below a square millimeter of cortical surface, as a universal
building block of brain-like computing. It is the smallest network
in which both a realistic number of $10\small,000$ synapses per neuron
and a connection probability of $0.1$ are realized simultaneously.

\begin{figure}
\centering{}\includegraphics[width=0.8\paperwidth]{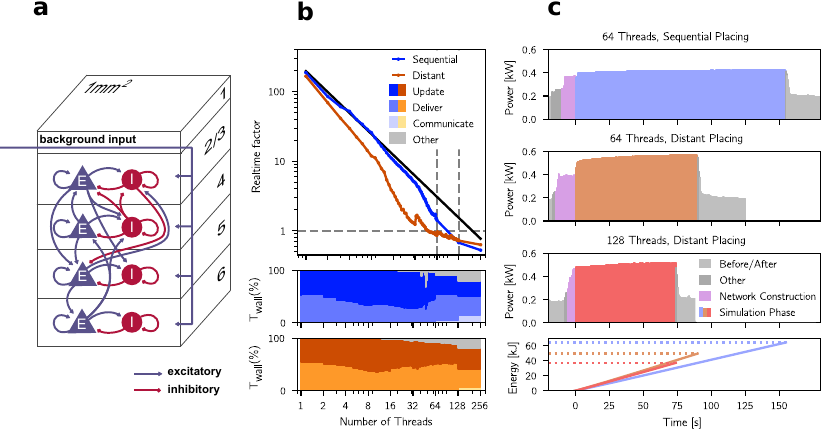}\caption{Strong scaling of a cortical microcircuit model on a conventional
compute node. \textbf{a }Sketch of the microcircuit model with about
$80,000$ neurons and $300$ million synapses organized into four
layers of excitatory (blue) and inhibitory (red) populations of neurons
(Supp. Inform. Fig. 1 shows activity). \textbf{b }Strong scaling for
two placing schemes. Top graph shows realtime factor over total number
of threads; dashed horizontal line indicates realtime; black solid
line indicates linear scaling. The sequential scheme (blue) minimizes
distance of threads on hardware, the distant scheme (brown) maximizes
it; dashed vertical lines indicate number of cores per processor ($64$)
and node ($128$). Bottom graphs show fractions of wall-clock time
consumed by different stages of the simulation cycle; update: integrates
state of neurons, deliver: distributes spike events to target neurons,
communicate: transfers spikes between MPI processes (for shared and
distributed memory setups), other: not accounted for by timers. \textbf{c
}Top three graphs: Power measurements of a compute node during $100\;\mathrm{s}$
of model time in three configurations. The measurements are aligned
to the start of the simulation phase starting at $t=0$ (legend: colors
distinguish phases and baseline). Bottom graph: Cumulative energy
consumption of the simulations.\label{fig:subreal1}}
\end{figure}
In a prototype network model of the microcircuit \citep{Potjans14_785},
the spatial structure of the cortex is neglected and replaced by cell-type
specific random connectivity. Each cortical layer is represented by
an excitatory and an inhibitory population of integrate-and-fire model
neurons (\prettyref{fig:subreal1}\textbf{a}).

The microcircuit model has become a benchmark for neuromorphic computing
systems: it can be simulated with moderate hardware investments \citep{VanAlbada18_291,Knight18_941},
its natural size renders questions of downscaling irrelevant \citep{Albada15},
and it marks an upper bound as larger neuronal networks are necessarily
less densely connected, and thus are, relative to the problem size,
easier to simulate.

Fast and energy efficient simulation is a promise of neuromorphic
computing \citep{Furber16_051001}; desirable for large-scale neuroscientific
models \citep{vanAlbada21_47} and imperative in artificial intelligence
and machine learning applications \citep{Strubell19}. The first milestone
is realtime performance, which was accomplished for the microcircuit
model in 2019 on a neuromorphic system \citep{Rhodes19_20190160}
followed this year by GPU systems \citep{Golosio21_627620,Knight21_15},
one of them already breaking into the sub-realtime regime \citep{Knight21_15}.
However, these results have to be evaluated in the light of continuously
advancing commodity hardware as a reference technology providing more
flexibility at potentially lower costs. With this aim we set out to
investigate the performance of the general purpose simulation engine
NEST \citep{Gewaltig_07_11204} on a recent conventional computing
system. Preliminary results have been presented in abstract form \citep{Kurth20_Bernstein}.

\section*{Methods }

We simulate the microcircuit model on $128$ core dual socket AMD
EPYC Rome 7702 compute nodes. Each processor is composed of $8$ chiplets,
each chiplet holds $8$ cores, resulting in $64$ cores per socket
(see Supp. Inform. Fig. 2). Each core has its own L1 and L2 cache,
$4$ cores share an L3 cache (see Supp. Inform. Fig. 3). Two nodes
are coupled by a point-to-point Mellanox ConnectX-6 HDR100 interconnect.
The software is NEST 2.14.1 \citep{Nest2141} (compiled with GCC 6.3.0
and using jemalloc 3.6.0-9.1 \citep{Evans06}, see Supp. Inform. Allocator)
providing, in contrast to some neuromorphic systems, double precision
numerics and weight resolution. NEST utilizes the Message Passage
Interface (MPI, here OpenMPI 4.0.3rc4 \citep{gabriel04_97}) and employs
hybrid parallelization with multithreading (OpenMP \citep{OpenMPSpec})
for shared memory parallelization where a core never runs more than
one thread. Timers monitor the different phases of the simulation.

Strong scaling experiments keep the task size fixed while systematically
increasing the computational resources (\prettyref{fig:subreal1}\textbf{b}).
The task is a simulation of $10\:\mathrm{s}$ of model time ($T_{\mathrm{{Model}}}$
), referring to the span of biological time described by the model,
if not stated otherwise. Measurements start after model instantiation
with optimized initial conditions \citep{Rhodes19_20190160} and an
initial interval of $\unit[0.1]{s}$ of model time to ensure that
potential transients of the network dynamics are discarded. To assess
simulation speed we use the realtime factor:
\[
\mathrm{RTF}=\frac{T_{\mathrm{Wall}}}{T_{\mathrm{Model}}}
\]
Here, $T_{\mathrm{Wall}}$ denotes the wall-clock time; the time passed
in the machine hall until the simulation completes. A realtime factor
smaller than $1$ implies sub-realtime performance. A common measure
for comparing the energy consumption of neuromorphic systems is energy
per synaptic event defined as total consumed energy divided by the
total number of transmitted spikes (see Supp. Inform. Power measurements).
For conducting the benchmarks we employ the JUBE \citep{Luehrs16_432}
benchmarking environment.

\section*{Results }

We assess the strong scaling performance of microcircuit model simulations
by increasing the number of threads on up to two compute nodes with
two different schemes of binding threads to cores: In the ``sequential''
placing scheme, threads are bound onto physically consecutive cores
per socket (thread counts $1$ to $64$ in steps of $1$), and $1$
MPI process per socket is used for simulations on one and two full
nodes with $128$ and $256$ threads, respectively. In the ``distant''
placing scheme, threads are bound such that L3 cache and chiplet overlap
is minimized per node (thread counts $1$ to $128$ in steps of $1$,
see Supp. Inform. Distant Placing) and $1$ MPI process per node is
used.

For sequential placing, we observe linear scaling for a thread count
between $1$ and $32$ as well as super-linear scaling between $32$
and $64$ (\prettyref{fig:subreal1}\textbf{b}). A full compute
node achieves sub-realtime performance with an RTF of $0.7$. Two
nodes reduce the realtime factor to $0.59$; the simulation runs $1.7$
times faster than realtime. The distant placing scheme exhibits super-linear
scaling already for a small number of threads. At $33$ threads, we
note a sudden rise of the realtime factor. At this point, the L3 cache
is shared for the first time. Nevertheless, sub-realtime performance
is already achieved when using only $64$ threads. Comparing the two
placings at $128$ and $256$ threads respectively, we observe that
sequential placing results in better performance. This is due to $2$
MPI processes being used on one node in the sequential placing scheme
as compared with $1$ for the distant placing. The relative time spent
in the update phase on a single node is decreased in the distant placing
when compared with the sequential one and communication between the
two nodes is not a limiting factor. This suggests that simulation
time can be further reduced by increasing the number of nodes and
alternatively using faster nodes.

We also assess the energy consumption of the simulation phase to investigate
how the increased power uptake due to using more computational resources
is counterbalanced by decreased simulation time (\prettyref{fig:subreal1}\textbf{c}).
For this we compare a configuration using all $128$ cores of a node
with two configurations using only half of the cores. The former sequentially
fills the cores of one socket, the latter employs the distant placing
scheme. During simulations of $\unit[100]{s}$ of model time we record
the power consumption and obtain the energy consumed in the simulation
phase by integrating over the power readings.

We observe that power consumption during the simulation phase is largest
for the distant placing of $64$ threads, amounting to $\unit[0.39]{kW}$
subtracting the baseline power of $\unit[0.2]{kW}$. This is almost
twice the power as in the sequential configuration ($\unit[0.21]{kW}$).
Nevertheless the increase cannot be attributed to the use of the second
socket. The $128$ thread configuration consumes $\unit[0.33]{kW}$
which is close to the same power required per thread of the sequential
case. The counterintuitively low power consumption in the $128$ threads
case may be explained by the potentially longer latencies resulting
in the cores not working at full capacity. Measuring the number of
cache misses confirms a relative frequency of $25\%$ in distant as
compared with $43\%$ in sequential placing (see Supp. Inform. Low
level performance measurements). Ultimately, the $128$ thread configuration
does not only exhibit the shortest time to solution but also requires
the smallest amount of energy.

The energy per synaptic event for the two fastest configurations ($128$
and $256$ threads in sequential placing) are $\unit[0.33]{\mu J}$
and $0.48\mu J$, respectively.

\section*{Discussion }

Our study shows that a single compute node achieves sub-realtime performance
in the simulation of a natural density local cortical microcircuit
model. To our best knowledge, we report the lowest realtime factor
so far at a competitive energy consumption(\prettyref{tab:comparison}).
There are, however, preliminary data \citep{Heittmann20_Bernstein}
on an even smaller realtime factor for a dedicated FPGA supercomputer
using on-the-fly generation of connectivity. Our results expose that
cache sensitive binding of threads increases performance.

\begin{table}
\begin{tabular}{|>{\centering}p{2cm}|c|l|}
\hline 
RTF & $E_{\mathrm{syn-event}}$ $\left(\mu\text{J}\right)$ & Reference\tabularnewline
\hline 
\hline 
$6.29$ & $4.39$ & 2018, NEST\citep{VanAlbada18_291}\tabularnewline
\hline 
$2.47$ & $9.35$ & 2018, NEST\citep{VanAlbada18_291}\tabularnewline
\hline 
$26.08$ & $0.30$ & 2018, GeNN\citep{Knight18_941}\tabularnewline
\hline 
$1.84$ & $0.47^{\dagger}$ & 2018, GeNN\citep{Knight18_941}\tabularnewline
\hline 
$1.00$ & $0.60$ & 2019, SpiNNaker\citep{Rhodes19_20190160}\tabularnewline
\hline 
$1.06$ & $-$ & 2021, NeuronGPU\citep{Golosio21_627620}\tabularnewline
\hline 
$0.70$ & $-$ & 2021, GeNN\citep{Knight21_15}\tabularnewline
\hline 
$0.67$ & $0.33$ & NEST, AMD EPYC Rome (single node)\tabularnewline
\hline 
$0.53$ & $0.48$ & NEST, AMD EPYC Rome (two nodes)\tabularnewline
\hline 
\end{tabular}
\centering{}\caption{\label{tab:comparison} Realtime factor (RTF) and energy per synaptic
event ($E_{\mathrm{syn-event}}$) reported in the literature  for
simulations of the cortical microcircuit model \citep{Potjans14_785}
using conventional hardware for NEST simulations, GPUs for GeNN and
NeuronGPU, and the dedicated neuromorphic hardware SpiNNaker in historical
sequence (top to bottom). The two values reported for NEST and GeNN
in 2018 (corresponding to the most energy efficient and the fastest
configuration) are obtained with a different number of employed cores
and different GPUs, respectively. $^{\dagger}$Value estimated by
the authors.}
\end{table}
Comparison with previous studies yields that conventional architectures
keep pace with dedicated hardware regarding both: realtime factor
and energy efficiency. The employed generic simulation engine for
spiking neuronal networks explicitly stores the connections between
neurons with double floating point precision. Thus, although not exploited
here, plasticity and learning are possible in this representation.
No attempt is made to optimize the simulation code for the particular
network model at hand. In comparison to prior work \citep{VanAlbada18_291},
where an earlier version (2.8.0) of the code and older hardware is
used, we observe a ten-fold improvement in performance. The older
system suffers from the communication between nodes as a bottleneck.
The newer hardware pushes the limits by integrating a larger number
of computational cores into the nodes. The analysis shows that on
a single node faster completion of the task comes with a lower energy
consumption due to the substantial baseline power. The simulation
time reduces if cores have a larger amount of cache available, and
if all cores are in use, power consumption is lower than for half
of the cores with optimal cache access. These observations indicate
that threads suffer from cache misses and the resulting latencies
in memory access. This does not only give practical guidance for the
design of conventional hardware but also raises hope that methods
of prefetching and latency hiding can further improve simulation code
without restricting generality\citep{pronold2021routing_efficient_cache}.

Achieving realtime performance is a criterion for robotics. But for
basic research and medical applications, also faster simulations are
of use, because biological processes extending over long periods of
time can be observed on a reduced time scale and multiple scenarios
can be investigated quickly.

We hope that our results further advance and inspire the constructive
competition between neuromorphic hardware and conventional computer
architectures \citep{VanAlbada18_291} which led to an order of magnitude
improvement within just four years.\ifthenelse{\boolean{isarxiv}}
{\section*{Data availability}
All data and analysis code to reproduce the results of this study can be downloaded from  \href{https://doi.org/10.5281/zenodo.5637375}{https://doi.org/10.5281/zenodo.5637375}.}{}
\begin{acknowledgments}
We are grateful to Tobias Noll and Arne Heittmann for fruitful discussions,
to Susanne Kunkel for comments on an earlier version of the manuscript,
to Jari Pronold for help with the JUBE code, to Sebastian Lehmann
for help with the design of the figures, to our colleagues in the
Simulation and Data Laboratory Neuroscience of the J\"ulich Supercomputing
Centre for continuous collaboration, and to the members of the NEST
development community for their contributions to the concepts and
implementation of the NEST simulator. Partly supported by the European
Union Seventh Framework Programme under grant agreement no. 604102
(Human Brain Project, HBP RUP), the European Union's Horizon 2020
(H2020) funding framework under grant agreement no. 720270 (HBP SGA1),
no. 785907 (HBP SGA2), no. 945539 (HBP SGA3), and the Helmholtz Association
Initiative and Networking Fund under project number SO-092 (Advanced
Computing Architectures, ACA). All network simulations were carried
out with NEST (\href{http://www.nest-simulator.org}{http://www.nest-simulator.org}).
\end{acknowledgments}

\bibliographystyle{\string~/texmf/bibtex/bib/iopart-num}

\providecommand{\newblock}{}

\end{document}

\section*{Supplementary Information to: Sub-realtime simulation of a neuronal
network of natural density}
\author{Anno C. Kurth$^{1,2}$, Johanna Senk$^{1}$, Dennis Terhorst$^{1}$,
Justin Finnerty$^{1}$ and Markus Diesmann$^{1,3,4}$}
\affiliation{$^{1}$Institute of Neuroscience and Medicine (INM-6) and Institute
for Advanced Simulation (IAS-6) and JARA-Institute Brain Structure-Function
Relationships (INM-10), Jülich Research Centre, Jülich, Germany~\\
$^{2}$RWTH Aachen University~\\
$^{3}$Department of Psychiatry, Psychotherapy and Psychosomatics,
School of Medicine, RWTH Aachen University, Aachen, Germany~\\
$^{4}$Department of Physics, Faculty 1, RWTH Aachen University, Aachen,
Germany}

\maketitle
\begin{figure}[H]
\begin{centering}
\includegraphics[width=12cm]{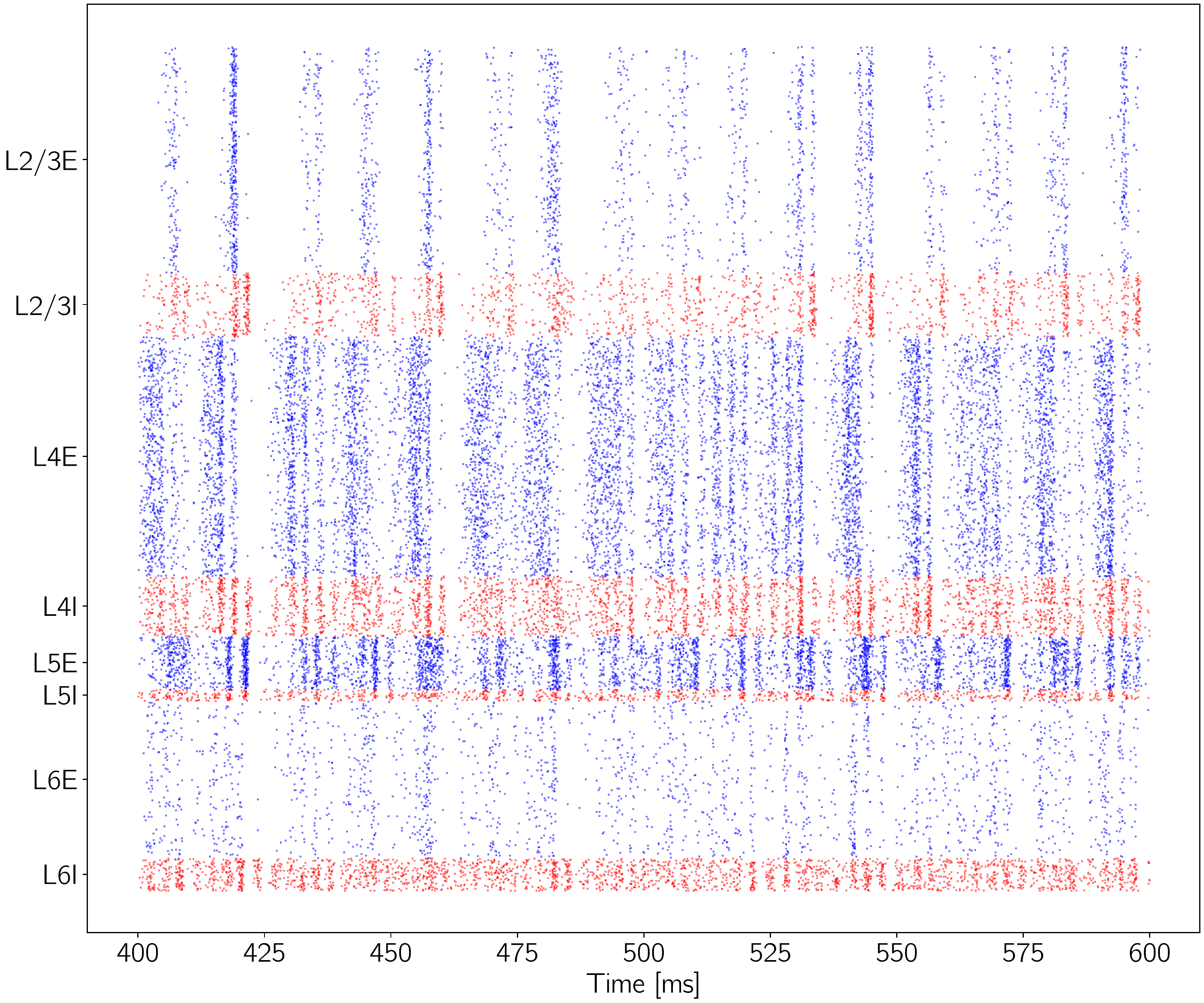}
\par\end{centering}
\caption{Raster plot of population resolved neuronal activity of a simulation
of the microcircuit model. For each population (vertical), the display
shows the spikes of a randomly selected fraction ($60\%$) of the
neurons in an arbitrary time segment of $200\:\mathrm{ms}$ (horizontal).
Blue dots correspond to excitatory, red dots to inhibitory neurons.
The temporal resolution of the simulation is $0.1\:\mathrm{ms}$,
the smallest delay in the network is $0.1\:\mathrm{ms}$, the smallest
time constant $\tau_{\mathrm{syn}}=0.5\:\mathrm{ms}$ (synaptic current),
the membrane time constant of the model neurons is $10\:\mathrm{ms}.$
The network exhibits spontaneous asynchronous irregular activity with
cell-type specific firing rates akin to experimental findings.}
\end{figure}
\begin{figure}[H]
\begin{centering}
\includegraphics[width=12cm]{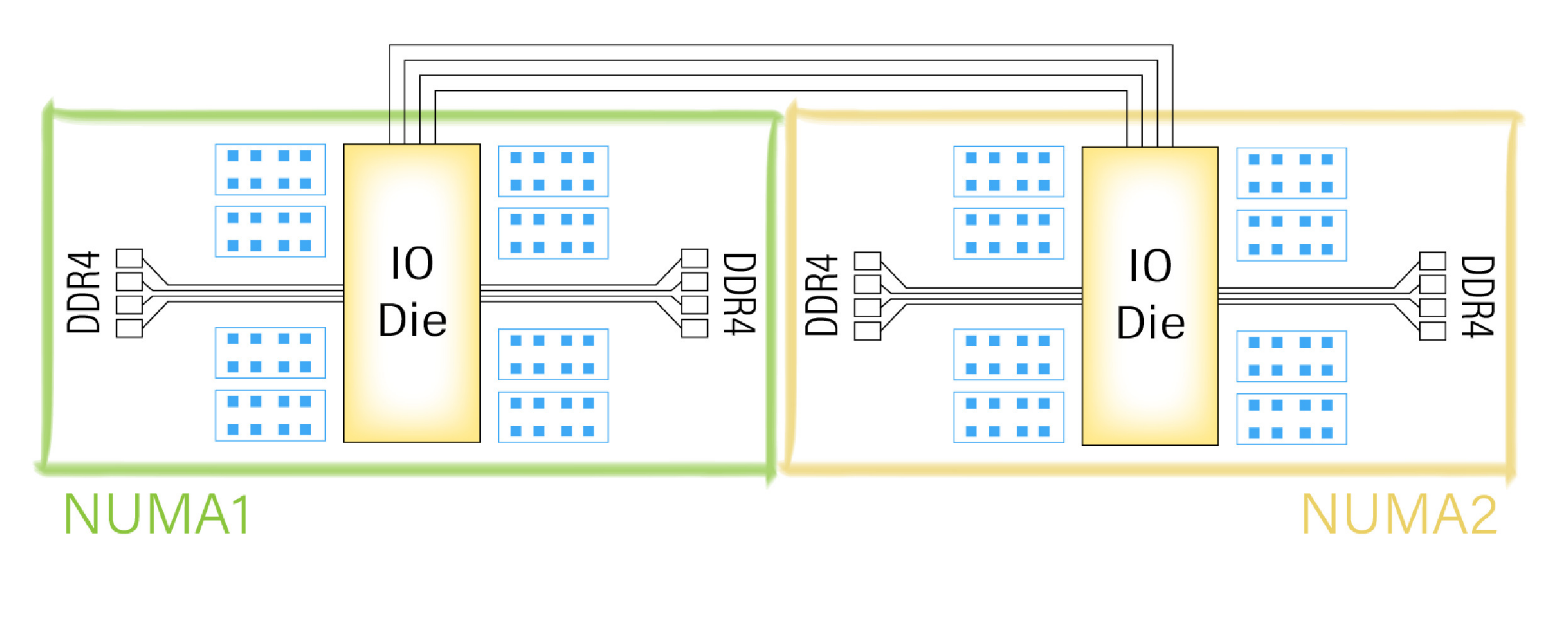}
\par\end{centering}
\caption{Sketch of hardware architecture of the dual socket AMD EPYC Rome 7702
system used in this study. Solid blue squares indicate compute cores.
$8$ compute cores are combined into one chiplet.\label{fig:hardware}}
\end{figure}
\begin{figure}[H]
\begin{centering}
\includegraphics[width=12cm]{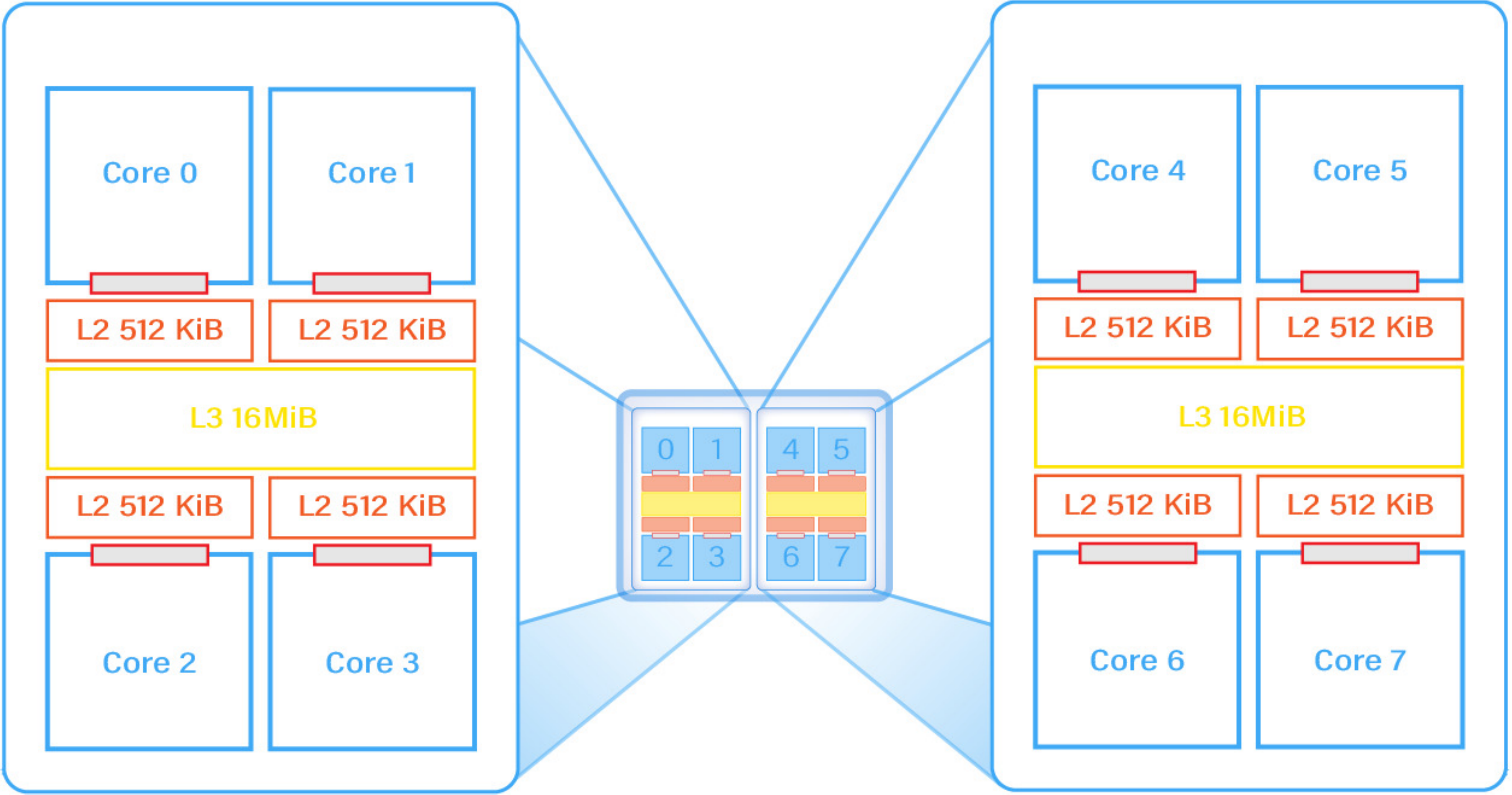}
\par\end{centering}
\caption{Sketch of one chiplet of the AMD EPYC Rome 7702. $4$ cores are grouped
into one core complex sharing an L3 cache.\label{fig:chiplet}}
\end{figure}

\subsection*{Allocator}

Using the preloading mechanism for shared libraries, we use the jemalloc
allocator via:

\begin{lstlisting}
export LD_PRELOAD=/path/to/libjemalloc.so.1 
\end{lstlisting}

\subsection*{Distant placing}

Let us first introduce a numbering scheme for the cores as sketched
in \Figref{hardware}. The kernel numbers the chiplets $0,...,15$
where $0,...,7$ identify consecutive chiplets on one socket and $8,...,15$
on the other. The numbering is induced by the standard output of $\texttt{lstopo}$
included in several Linux distributions. The command returns a numbered
list of the cores on the compute node hierarchically structured by
the NUMA nodes (in our case equivalent to the sockets), the L3 cache
and the L1/L2 cache. Cores $0$ to $63$ and L3 caches $0$ to $15$
are located on NUMA node $0$, cores $64$ to $127$ and L3 caches
$16$ to $31$ on NUMA node $1$. Since on one chiplet two L3 caches
are located, one obtains the number of a chiplet by an integer division
of the number of the respective L3 cache by $2$. We denote the $k\mathrm{-th}$
core, $k\in\{0,...,7\}$ (sketched in \Figref{chiplet}), on the $n\mathrm{-th}$
chiplet by $n:k$. In the distant placing scheme the filling of a
compute node with threads is split into $8$ rounds each addressing
a particular core $k$ of the chiplets and successively adding this
core of chiplet $n$ ($16$ in total) to the simulation. This results
in $8\times16=128$ threads being bound to cores. The filling procedure
starts with core $0$, i.e. the first $16$ use the cores $\{0:0\},\{0:0,1:0\},...,\{0:0,...,15:0\}$.
The next round uses cores still not sharing an L3 cache with cores
already in use. We chose the $4\mathrm{-th}$, resulting in consecutively
adding the cores $0:4,1:4,...,15:4$. The following rounds continue
with the $2\mathrm{-nd}$, $6\mathrm{-th}$, $1\mathrm{-st}$, $5\mathrm{-th}$,
$3\mathrm{-rd}$ and $7\mathrm{-th}$ core respectively, minimizing
shared use of L3 cache. To bind threads to cores on our system we
export OpenMP variables as follows:
\begin{lstlisting}
export OMP_NUM_THREAD = $CPUSPERTASK
export OMP_PROC_BIN = TRUE
export OMP_PLACES = {0},{8},{15}
\end{lstlisting}
Here $\texttt{\$CPUSPERTASK}$ is the number of cores used in a given
setup (in this example $3$) and $\texttt{\{0\},\{8\},\{15\}}$ indicate
the first core on the first, second and third chiplet. Simulations
on one node are launched by

\begin{lstlisting}
python3 run_microcircuit.py  
\end{lstlisting}

Simulation on two nodes are launched by

\begin{lstlisting}[breaklines=true]
mpirun --n 2 --npernode 1 --mca pml ucx -x UCX_NET_DEVICES=mlx5_1:1 --bind-to board python3 run_microcircuit.py 
\end{lstlisting}
in this example with $1$MPI process per node.

\subsection*{Power measurements}

Power was measured with a Raritan Dominion PX and a Raritan PX3-5190
power distribution unit (PDU). The units have an accuracy of $\pm5\%$
and data collection frequency of $1\;\mathrm{Hz}$. The power measurement
has a delay of $1\:\mathrm{s}$, so that the power readings need to
be shifted by $1\;\mathrm{s}$ to be aligned to wall-clock time. Since
the nodes are connected point-to-point, we do not need to take additional
passive energy consumption by an interconnect into account.

\subsection*{Low level performance measurements}

\noindent In order to determine the number of cache misses we employ
the $\texttt{\texttt{perf}}$ performance analysis tool of the Linux
operating system. We use the command options

\noindent 
\begin{lstlisting}[breaklines=true]
perf stat -ae task-clock,cycles,instructions,cache-references,cache-misses
\end{lstlisting}
 and increase the simulation time to $100\;\mathrm{s}$. With this
we ensure that approximately $80\%$ of the run time of the program
is spent in the simulation phase guaranteeing a reliable assessment
of the percentage of cache misses during that phase.\ifthenelse{\boolean{isarxiv}}
{}{\section*{Data availability}
All data and analysis code to reproduce the results of this study can be downloaded from  \href{https://doi.org/10.5281/zenodo.5637375}{https://doi.org/10.5281/zenodo.5637375}.}